\documentstyle[array,epsfig,twocolumn,pra,aps]{revtex}
\begin{document}
\draft

\wideabs{
  
  \title{Optimal conclusive teleportation of a $d$-dimensional unknown
    state}
  
  \author{W. Son,$^1$\footnote{sonwm@family.sogang.ac.kr} Jinhyoung
    Lee,$^1$\footnote{hyoung@quanta.sogang.ac.kr} M. S.
    Kim$^2$\footnote{m.s.kim@qub.ac.uk}, and Y.-J.
    Park$^1$\footnote{yjpark@ccs.sogang.ac.kr}}
  
  \address{$^1$ Department of Physics, Sogang University,
    CPO Box 1142, Seoul 100-611, Korea \\
    $^2$School of Mathematics and Physics, The Queen's University,
    Belfast BT7 1NN, United Kingdom}
  
  \date{\today}
  
  \maketitle

\begin{abstract}
  We formulate a conclusive teleportation protocol for a system in
  $d$-dimensional Hilbert space utilizing the positive operator valued
  measurement at the sending station.  The conclusive teleportation
  protocol ensures some perfect teleportation events when the channel
  is only partially entangled, at the expense of lowering the overall
  average fidelity.  We find the change of the fidelity as optimizing
  the conclusive teleportation events and discuss how much information
  remains in the inconclusive parts of the teleportation.
\end{abstract}

}


\section{INTRODUCTION}
\label{sec:sec1}

Quantum teleportation \cite{Bennett93,Zeilinger97} serves probably the
best test ground for quantum entanglement.  When its quantum channel
is optimally entangled, quantum teleportation faithfully transmits the
quantum state of a system.  It has been found for a spin-1/2 system
that even when the quantum channel is not maximally entangled, quantum
teleportation provides better transmission of a quantum state than any
classical communication protocol \cite{Popescu94}.

When an imperfect quantum channel is given, there are two components
we can alter in quantum teleportation to satisfy the needs: {\it
  measurement} at the sending station and {\it transformation} at the
receiving station.  Banaszek found that, when the channel is {\it
  pure}, a joint von-Neumann measurement and a unitary transformation
maximizes the average fidelity for the teleportation of a state in
$d$-dimensional Hilbert space \cite{Banaszek00}.  In some cases, this
may not be the optimal strategy for particular purposes.  For example,
when a perfect replica of the original state is only of use,
maximizing the average fidelity does not have to be the best strategy
because it can mean none of the replica states are perfect even though
the density matrix of the replica ensemble maximally resembles the
original state.  In this paper we are interested in how to optimize
the successful teleportation of a quantum state in $d$-dimensional
Hilbert space.

Recently, Mor and Horodecki \cite{Mor99} proposed a conclusive
teleportation protocol for a quantum state in two-dimensional Hilbert
space by employing a positive operator valued measurement (POVM)
\cite{Jauch67} as the joint measurement at the sending station.  When
the measurement is successful by a random chance, the initially
unknown quantum state is teleported perfectly, which is called a {\it
  conclusive} event. In contrast, for an {\it inconclusive} event, the
sender can extract no information from the measurement outcome and the
teleportation loses its quantum characteristics.  To indicate the
success of the measurement, an extra single-bit has to be sent to the
receiver together with the two-bit information via a classical
channel.  Bandyopadhyay \cite{Bandyopadhyay00} and Li {\it et al.}
\cite{Li00} have also proposed protocols to implement conclusive
teleportation in two-dimensional Hilbert space.  Bandyopadhyay uses a
combination of orthogonal GHZ measurements and POVM's while Li {\it et
  al.}  take general transformations at the receiving station leaving
the measurement orthogonal.

Quantum information theory has been extensively developed for a
two-level spin-1/2 system because this simple model provides a natural
extension of a binomial bit to a quantum bit, namely, {\it qubit}, and
gives a comprehensive understanding of fundamental quantum theory.  It
is only recently that $d$-dimensional quantum systems have attracted a
considerable research effort.  Quantum cloning in $d$ dimensions was
studied by Zanardi \cite{Zanardi98} and quantum teleportation by
Zubairy \cite{Zubairy98} and Stenholm and Bardroff \cite{Stenholm98}.
Rungta {\it et al.} called the $d$-dimensional quantum system the {\it
  qudit} and investigated its entanglement and separability
\cite{Rungta00}.  The qudit was extensively studied as a
finite-dimensional version of a continuous variable state by Gottesman
{\it et al.}  \cite{Gottesman00}.

The POVM is not a new concept in quantum information theory.  The POVM
is one of the standard procedures for entanglement purification and
concentration.  The entanglement purification is a protocol to select
the subensemble of maximally entangled pairs from the whole ensemble
of quantum states \cite{Bennett96}.  In the purification protocol, two
distant observers employ only local measurements and classical
communication together with the post-selection of subensembles. The
local measurements may involve POVM's \cite{Vedral98}.  The
teleportation procedure may follow the purification of the quantum
channel for the faithful transmission of a quantum system.  The
optimal POVM has recently been implemented in a laboratory for a
two-level system \cite{Clarke00}.

In this paper, we formulate the conclusive teleportation in
$d$-dimensional Hilbert space, for which we develop the general form
of the optimum POVM. Here, the conclusive measurement outcome is not
to give information on the initial state itself but to give
information on which unitary transformation to perform to recover the
initial state at the receiving station.  An optimum POVM normally
leaves no information in the inconclusive event but when the POVM is
used in teleportation, we show that some teleportation information can
be extracted from the inconclusive outcome.  However, in this case,
the maximum information the receiver recovers is limited by classical
theory.  If the POVM is not optimized, some {\it quantum} information
is still left in the inconclusive event.  We show how much quantum
information remains in the inconclusive event by evaluating the
teleportation fidelity.  The overall fidelity for conclusive
teleportation is less than that for the teleportation employing the
von-Neumann measurement.  We closely examine the reason.  We also
assess the minimum resources to achieve the conclusive teleportation.

\section{conclusive teleportation}
\label{sec:sec2}

Conclusive teleportation assures some faithful teleportation events
using a partially-entangled quantum channel.  In this section, we
formulate conclusive teleportation in the $d$-dimensional Hilbert
space $\cal H$. One of the important conditions for teleportation is
that the initial quantum state is unknwon.  Assume that a particle in
a $d$-dimensional quantum state $| \phi\rangle_1$,
\begin{equation}
\label{eq:unknown}
|\phi\rangle_1 = \sum^d_{i=1} c_i | i\rangle_1,
\end{equation}
is teleported to a remote place via a partially entangled channel in
the $d\times d$-dimensional Hilbert space, $\cal H\otimes\cal H$.

When the quantum channel is not perfect, Mor and Horodecki
\cite{Mor99} found for a two-level quantum system that a joint POVM
may assure conclusive teleportation.  The main difference between a
POVM and a von-Neumann orthogonal measurement is that the outcome
bases and the number of outcomes may be different each other
\cite{Peres93}.  By discerning non-orthogonal states using a joint
POVM, a faithful transmission of a two-level quantum state is made
possible.  Of course, it is not always possible to discern
non-orthogonal states due to the overlap between the non-orthogonal
outcome bases.  For the inconclusive outcome, the teleportation
becomes unfaithful.  In any cases, we know when the teleportation is
faithful.  We formulate the optimal conclusive teleportation for a
quantum state in $d$-dimensional Hilbert space.

\subsection{Formalism}
\label{sec:sec2i}

For quantum teleportation, we assume that the quantum channel is
prepared with a {\it pure entangled pair} of particles 2 and 3.  The
quantum channel is in the state $| \psi\rangle_{23}$ by Schmidt
decomposition
\begin{equation}
\label{eq:initial channel}
|\psi \rangle_{23} = \sum ^{d} _{i=1} a_i |i i \rangle_{23}
\end{equation}
where $\{| i\rangle\}$ is an orthonormal basis set in $d$-dimensional
Hilbert space $\cal H$.  The coefficients $a_i$ are regarded as real
values and assume that $a_k$ is the smallest among the coefficients
for the sake of simplicity. For a maximally entangled quantum channel,
$a_i =d^{-1/2}$. The total quantum state for the unknown particle 1
and the entangled pair 2 and 3 is given by the direct product of the
unknown state $|\phi\rangle_1$ and the quantum channel state
$|\psi\rangle_{23}$,
\begin{equation}
\label{eq:initial}
| \Psi\rangle_{123}=| \phi\rangle_1 \otimes | \psi
\rangle_{23}.
\end{equation}
The sender performs a joint measurement on particles 1 and 2 so that
we expand the composite system based on states for particles 1 and 2.

In the standard teleportation protocol suggested by Bennett {\it et
  al.} \cite{Bennett93}, the joint measurement is based on Bell
states, which form an orthonormal basis set of the maximally entangled
states $\{|\psi^m_{\alpha}\rangle\}$ for a spin-1/2 system.  This
basis set can be obtained by applying a set of local unitary
operations $\{\hat{U}^{\alpha}\}$ on the given maximally entangled
state $|\psi^m \rangle$: $|\psi^m_{\alpha}\rangle= \hat{U}^{\alpha}
\otimes \openone |\psi^m \rangle$ \cite{Braunstein00}.  For a spin-1/2
system the set of unitary operators is given by $\{\openone,
\hat{\sigma}_x, \hat{\sigma}_y, \hat{\sigma}_z\}$. The orthonormal
basis set should satisfy the completeness:
\begin{eqnarray}
  \label{eq:cr}
  \openone_{d \times d} &=&\sum_{\alpha=1}^{d^2}
  |\psi^m_{\alpha}\rangle \langle \psi^m_{\alpha}|  \nonumber \\ 
  &=& \sum_\alpha \hat{U}^{\alpha} \otimes \openone |\psi^m
  \rangle \langle \psi^m | \hat{U}^{\alpha^\dagger}\otimes\openone,
\end{eqnarray}
where $\openone_{d \times d} = \openone \otimes \openone$.  In the
orthonormal basis of $\{|ij \rangle \}$ the completeness can be
written in the matrix form:
\begin{equation}
  \label{eq:crmf}
  \delta_{ik}\delta_{jl} = \frac{1}{d} \sum_{\alpha} U^{\alpha}_{ij}
  U^{\alpha^*}_{kl}
\end{equation}
where $U^\alpha_{ij} \equiv \langle i |\hat{U}^\alpha|j \rangle$. Note
that Eq.~(\ref{eq:crmf}) depends only on the unitary operators. In
fact the set of the unitary operators $\{\hat{U}^\alpha\}$ is the {\it
  irreducible projective representation} of a group $G$ and
Eq.~(\ref{eq:crmf}) implies its orthogonality relation
\cite{Altmann77}.  In addition the orthonormality of the entangled
basis states $\delta_{\alpha\beta}=\langle\psi^m_\alpha|\psi^m_\beta
\rangle$ leads to the orthogonal condition for the unitary operators:
\begin{eqnarray}
  \label{eq:traceless}
  \mbox{Tr} \hat{U}^{\alpha^\dagger} \hat{U}^\beta = d~
  \delta_{\alpha\beta}.
\end{eqnarray}

In the conclusive teleportation protocol, the basis set of entangled
states is obtained by using the same local unitary operators
$\{\hat{U}^\alpha\}$ on the state of the quantum channel
(\ref{eq:initial channel}):
\begin{equation}
\label{eq:set of basis}
|\psi_{\alpha}\rangle = \hat{U}^{\alpha} \otimes \openone |\psi
\rangle
\end{equation}
for $\alpha=1,2,\cdot\cdot\cdot,d^2$. Note that the state vectors
(\ref{eq:set of basis}) are linearly independent and form a basis set.
The basis states $|\psi_{\alpha}\rangle$ are not necessarily
orthonormal.  Only when the channel is maximally entangled with
$a_i=d^{-1/2}$ the set of the basis states $\{|\psi_{\alpha}\rangle\}$
becomes $\{|\psi_{\alpha}^m\rangle\}$, which represents the
von-Neumann orthogonal measurement.

It is possible to write the state $|\psi_{\alpha}\rangle$ in the
orthonormal basis $|ij\rangle$:
\begin{equation}
\label{eq:gamma1}
| \psi_{\alpha} \rangle = \sum^{d}_{i,j=1} \Gamma^{\alpha}_{ij} | ij
\rangle 
\end{equation}
where the matrix $\Gamma^{\alpha}_{ij}$ is calculated from
Eqs.~(\ref{eq:initial channel}) and (\ref{eq:set of basis}) as
\begin{equation}
\label{eq:gamma}
\Gamma^{\alpha}_{ij}=U^{\alpha}_{ij} a_j.
\end{equation}
Because the basis states $\{|\psi_{\alpha}\rangle\}$ are linearly
independent, the matrix $\Gamma^{\alpha}_{ij}$ is non-singular and its
inverse matrix is given by
\begin{equation}
\label{eq:gamma2}
\left(\Gamma^{-1}\right)^{\alpha}_{ij} = \frac{1}{d}
U^{\alpha^*}_{ij}a^{-1}_{j}. 
\end{equation}
It is straightforward to show, from Eqs.~(\ref{eq:gamma}) and
(\ref{eq:gamma2}) by using Eq.~(\ref{eq:crmf}) and/or
Eq.~(\ref{eq:traceless}), that Eq.~(\ref{eq:gamma2}) is the inverse
matrix for $\Gamma^{\alpha}_{ij}$ such that $\sum_\alpha
(\Gamma^{-1})^{\alpha}_{ij} \Gamma^{\alpha}_{kl}=
\delta_{ik}\delta_{jl}$ and/or $\sum_{ij} \Gamma^{\alpha}_{ij}
(\Gamma^{-1})^{\beta}_{ij} = \delta_{\alpha\beta}$.  With the help of
$(\Gamma^{-1})^{\alpha}_{ij}$, the inverse relation to
Eq.~(\ref{eq:gamma1}) follows
\begin{equation}
\label{eq:inverse}
|ij\rangle = \sum^{d^2}_{\alpha=1}
\left(\Gamma^{-1}\right)^{\alpha}_{ij} |\psi_{\alpha}\rangle.  
\end{equation}
The completeness with respect to the entangled states
$\{|\psi_{\alpha}\rangle\}$ is nontrivial due to their being
non-orthogonal for a partially entangled quantum channel. Instead, we
modify the completeness for the set of orthonormal bases $\{| ij
\rangle\}$ using Eq.~(\ref{eq:inverse}) as
\begin{eqnarray}
\label{eq:modcomplete}
\openone_{d \times d} &=&\sum_{ij} |ij\rangle\langle
ij| \nonumber\\ 
&=&\sum _{\alpha ij} \left(\Gamma^{-1}\right)^{\alpha}_{ij} | \psi_{\alpha}
\rangle \langle ij |.
\end{eqnarray}

The total state $|\Psi\rangle_{123}$ in Eq.~(\ref{eq:initial}) can now
be written with help of the modified completeness
(\ref{eq:modcomplete}) as
\begin{eqnarray}
\label{eq:Psi}
| \Psi\rangle_{123}
&=&\left[\sum _{\alpha ij} \left(\Gamma^{-1}\right)^{\alpha}_{ij} |
  \psi_{\alpha} 
  \rangle_{12}\langle ij |\right]| \phi\rangle_1 \otimes | \psi
\rangle_{23} \nonumber \\
&=& \frac{1}{d} \sum_{\alpha} |\psi_{\alpha}\rangle_{12} \otimes
\hat{U}^{\alpha\dagger} |\phi\rangle_3.
\end{eqnarray}
The second equality is given by the inverse matrix (\ref{eq:gamma2})
and the orthonormality of the bases $\{|i\rangle\}$.  Here it is seen
that the von-Neumann orthogonal measurement with the maximally
entangled bases of $\{|\psi^m_{\alpha}\rangle\}$ cannot exactly
discern the non-orthogonal state vectors $|\psi_{\alpha}\rangle$ and
the teleportation is no longer perfect for the partially entangled
quantum channel.

Suppose that there were a complete set of measurement operators
$\{\hat{A}_\alpha\}$ with $d^2$ outcomes, which could identify a
partially entangled state $|\psi_\beta\rangle$ such that the
probability $p(\alpha|\beta)=\langle \psi_\beta| \hat{A}_\alpha
|\psi_\beta\rangle$ does not vanish only when $\alpha=\beta$. One
could then achieve perfect teleportation by applying a unitary
operation $\hat{U}_\beta$ on particle 3, accordingly. Except for the
maximally entangled quantum channel, there does not exist such a set
of measurement operators $\{\hat{A}_\alpha\}$.  We thus adopt a set of
POVM operators $\{\hat{M}_\alpha\}$ with $n > d^2$ outcomes, which are
designed to satisfy the above condition of $p(\alpha|\beta)=\langle
\psi_\beta| \hat{M}_\alpha |\psi_\beta\rangle \propto
\delta_{\alpha\beta}$ for $\alpha \le d^2$ and to allow that
$p(\alpha|\beta)$ does not have to vanish for $d^2 < \alpha \le n$.
The outcomes are classified into two events. One is {\it conclusive}
for the outcomes of $\alpha \le d^2$ where we can achieve a perfect
transfer of the unknown state.  The other is {\it inconclusive} for
$d^2 < \alpha \le n$ where the measurement does not tell its outcome
precisely.  In the conclusive teleportation, the joint POVM determines
whether the present event is conclusive.  If the measurement outcome
is conclusive, a corresponding unitary operation completes the
teleportation.

\subsection{Identification by POVM}
\label{sec:sec2ii}

A POVM is defined as a partition of unity by the nonnegative operators
which are in general non-orthogonal. A set of POVM operators
$\{\hat{M}_{\alpha}\}$ with $n > d^2$ outcomes satisfy the measurement
conditions of positivity and completeness.  The positivity
$\hat{M}_{\alpha} \ge 0$ ensures the positive probability for every
POVM operator.  The fact that the sum of probabilities is unity leads
to the completeness: $\sum_{\alpha=1}^n \hat{M}_{\alpha} =\openone_{d
  \times d}$.

The conclusive teleportation has a crucial step to identify
non-orthogonal basis states $|\psi_{\beta}\rangle$ for
$\beta=1,2,...,d^2$ as in Eq.~(\ref{eq:Psi}).  For this purpose, joint
POVM operators are designed such that
\begin{eqnarray}
\label{eq:identify}
\langle \psi_\beta | \hat{M}_{\alpha}| \psi_{\beta} \rangle &\propto&
\delta_{\alpha\beta} ~~~~\mbox{for}~~~\alpha \le d^2.
\end{eqnarray}
This shows that when the measurement outcome is due to any of
$\hat{M}_{\alpha}$ for $\alpha \le d^2$, we can conclusively determine
which non-orthogonal state the system is in. On the other hand, the
measurement bears an inconclusive result when the outcome is of
$\hat{M}_{\alpha}$ for $\alpha > d^2$. In the conclusive
teleportation, $d^2$ operators are used to identify non-orthogonal
states while the other operators consisting of the measurement set
represent some inconclusive mixtures.

Any set of POVM operators can be decomposed into rank-one general
projectors \cite{Peres93}.  The conclusive measurement operators are
represented by general projectors as
\begin{equation}
\label{eq:operator}
\hat{M}_{\alpha} = \lambda_\alpha | \tilde{\psi}_{\alpha} \rangle\langle
\tilde{\psi}_{\alpha} | ~~~~~\mbox{for}~~\alpha \le d^2
\end{equation}  
where the real parameter $\lambda_\alpha \ge 0$ will be determined to
optimize the conclusive events while keeping the probability of the
inconclusive events positive.  Without loss of generality we assume
that each measurement outcome is equally probable with
$\lambda_\alpha=\lambda$. Unless the POVM is to discern orthogonal
states, the completeness is guaranteed only by adding an inconclusive
measurement operator $\hat{M}_{d^2+1}$,
\begin{equation}
  \label{eq:last operator}
  \hat{M}_{d^2+1}=\openone_{d \times d} - \sum_{\alpha=1}^{d^2}
  \hat{M}_\alpha. 
\end{equation}

For the purpose of the identification (\ref{eq:identify}), the
generally non-orthogonal and unnormalized states
$\{|\tilde{\psi}_{\alpha} \rangle\}$ in Eq.~(\ref{eq:operator}) are
constrained to satisfy the relation
\begin{equation}
\label{eq:ortho}
\langle \tilde{\psi}_{\alpha}|\psi_{\beta}\rangle =
\delta_{\alpha\beta}. 
\end{equation}
To find its explicit form, we expand $| \tilde{\psi}_{\alpha} \rangle$
in the orthogonal basis $|ij\rangle$,
\begin{equation} 
\label{eq:gamma4}
| \tilde{\psi}_{\alpha} \rangle = \sum^{d}_{i,j=1}
  \tilde{\Gamma}^{\alpha}_{ij} | ij \rangle,
\end{equation}
where $\tilde{\Gamma}^{\alpha}$ is given by the relation between
$\Gamma$ and $\Gamma^{-1}$ in Eqs.~(\ref{eq:gamma}) and
(\ref{eq:gamma2}) as
\begin{equation}
\label{eq:gamma5}
\tilde{\Gamma}^{\alpha}_{ij} = \left(\Gamma^{-1*}\right)^{\alpha}_{ij} 
= \frac{1}{d}U^{\alpha}_{ij} a^{-1}_j.
\end{equation}
It is straightforward to show that the unnormalized states
${|\tilde{\psi}_{\alpha} \rangle}$ satisfy the orthogonal condition
(\ref{eq:ortho}).

The inconclusive measurement operator $\hat{M}_{d^2+1}$ is determined
to satisfy the positivity and the condition (\ref{eq:last operator}).
To do so the sum of the conclusive measurement operators is calculated
as
\begin{eqnarray}
  \label{eq:complete}
  \sum^{d^2}_{\alpha=1} \hat{M}_{\alpha} =
  \frac{\lambda}{d^2}\sum_{ijkl} \frac{1}{a_j a_l} \left( \sum_{\alpha}
    U^{\alpha}_{ij} U^{*\alpha}_{kl}\right) |ij\rangle\langle kl|
\end{eqnarray}
With use of the orthogonality of unitary operators (\ref{eq:crmf}),
the inconclusive measurement operator is then found as
\begin{eqnarray}
\label{eq:last}
\hat{M}_{d^2+1} &=&\sum^{d}_{i,j=1} \left(1-\frac{\lambda}{
    a^2_j d}\right)|ij\rangle\langle ij|.
\end{eqnarray}
This is positive only when $\lambda \le a_j^2 d$ for all
$j=1,2,...,d$.  As $a_k$ the smallest, the condition, $0 \le\lambda
\le a_k^2 d$, ensures the positivity.  The operator $\hat{M}_{d^2+1}$
is diagonal and a convex combination of projectors.  Note that it is
possible to decompose the operator $\hat{M}_{d^2+1}$ further into
$d^2$ general projectors.  We then have a new set of POVM operators,
{\it i.e.}, $\{\hat{M}_1,\hat{M}_2,\cdot\cdot\cdot,\hat{M}_{2d^2} \}$.

\subsection{Optimization of conclusive events}
\label{sec:sec2iii}

The optimization of conclusive-event probabilities is achieved by
minimizing the probability of the inconclusive event,
\begin{equation}
  \label{eq:inconevent}
  p_{d^2+1} ={_{123}\langle}\Psi|\hat{M}_{d^2+1}|\Psi\rangle_{123}=
  1-\lambda, 
\end{equation} 
which shows that the maximum possible value of $\lambda$ will result
in the optimal conclusive teleportation. From the positivity condition
for all POVM operators, it is clear that $\lambda$ has the maximum
value, $a^2_k d$.  Note that, with the information on the channel, it
is always possible to perform a POVM which optimizes faithful
teleportation.

The proposed POVM enables to identify non-orthogonal states
$|\psi_{\alpha}\rangle$ with finite probability $p_\alpha=\lambda/d^2$
for each conclusive event. When it is employed for the joint
measurement, we have a nonzero probability to teleport faithfully.
When the measurement outcome is inconclusive, we simply repeat the
protocol till a conclusive result is obtained.

\subsection{Necessary resources}
\label{sec:2iv}

Our protocol for the conclusive teleportation has two distinct
components from standard teleportation: POVM for joint measurement and
additional classical communication whether the event is conclusive or
not. The additional classical communication requires single classical
bit.  Here, we consider what other resources are required to implement
conclusive teleportation.

Neumark's theorem enables us to construct the POVM by von-Neumann
orthogonal measurement in an extended Hilbert space. The extension is
done by adding an ancillary particle \cite{Peres93}.  Let $d_a$ the
dimension of the ancillary particle. In the extended Hilbert space
${\cal H}_s \otimes {\cal H}_a$, where ${\cal H}_s = {\cal
  H}\otimes{\cal H}$ is the original Hilbert space and ${\cal H}_a$
the ancillary Hilbert space, the von-Neumann orthogonal measurement is
represented by a set of projectors $\{\hat{P}_{ij} \otimes
\hat{P}_a\}$ satisfying the completeness,
\begin{equation}
  \label{eq:eHcr}
  \sum_{ija}\hat{P}_{ij} \otimes \hat{P}_a =
  \openone_{d \times d} \otimes \openone_{d_a}
\end{equation}
where $\hat{P}_{ij}=|ij\rangle \langle ij|$ and $\hat{P}_a=|a\rangle
\langle a|$ are projectors, respectively, in ${\cal H}_s$ and ${\cal
  H}_a$.  The set of POVM operators (\ref{eq:operator}) and
(\ref{eq:last operator}) for the joint measurement are constructed by
applying unitary operation $\hat{U}$ on the composite system of the
original system and the ancillary system and by projecting into the
original Hilbert space as
\begin{eqnarray}
  \label{eq:ptos}
  \openone_{d \times d} &=& \hat{P}
  \hat{U} \sum_{ija} |ij\rangle \langle ij| \otimes |a\rangle
  \langle a| \hat{U}^\dagger \hat{P}
\end{eqnarray}
where $\hat{P}$ is a projector into ${\cal H}_s$ such that $\hat{P}
|ij\rangle \otimes |a\rangle = |ij\rangle$. Now, the set of operators
$\{\hat{P}\hat{U} |ij\rangle \langle ij| \otimes |a\rangle \langle a|
\hat{U}^\dagger \hat{P}\}$ gives the set of POVM operators
$\{\hat{M}_\alpha\}$ if the following $d^2$ matrix equations are
satisfied,
\begin{equation}
  \label{eq:lets}
  \left(\hat{M}_\alpha\right)_{kl,mn} = \sum_{ija}
  \left(\hat{P}\hat{U}\right)_{kl,ija}
  \left(\hat{U}^\dagger\hat{P}\right)_{ija,mn} 
\end{equation}
for $\alpha \le d^2$, where $(\hat{M}_\alpha)_{kl,mn} = \langle kl|
\hat{M}_\alpha |mn\rangle$ and $(\hat{P}\hat{U})_{kl,ija} = \langle
kl|\hat{P}\hat{U}|ija\rangle$. Note that we consider only $d^2$ matrix
equations for the POVM operators used in conclusive events because
others are straightforwardly obtained from the completeness condition
(\ref{eq:ptos}).  The unitary operator has $(d \times d \times d_a)^2$
independent real variables and Eq. (\ref{eq:lets}) has $d^2 \times
(d^2 \times d^2)$ linear equations. If $d_a \ge d$, the number of real
variables are sufficient to satisfy Eq. (\ref{eq:lets}). Therefore,
the joint POVM requires an ancillary particle in $d_a \ge d$
dimensional Hilbert space.

\section{Average Fidelity}
\label{sec:sec3}

For a maximally entangled quantum channel the proposed conclusive
teleportation becomes the standard teleportation as the joint
measurement performed at the sending station becomes the orthogonal
measurement. In this case the joint measurement extracts no
information on the unknown quantum state as the full information is
transferred to the teleported state.  On the other hand, for a
partially entangled quantum channel, the joint orthogonal measurement
may extract some information on the unknown state and only partial
information is transferred to the teleported state.  The average
fidelity, which tells how close the teleported state is to the
original state, is not larger than $\bar{\cal F}_{s} =
\frac{2}{3}(1+|a_1a_2|)$ for the quantum teleportation of a
two-dimensional state using the standard teleportation protocol
\cite{Popescu94,Banaszek00,Horodecki96}.  Conclusive teleportation
employs the joint POVM instead of the orthogonal measurement. The
measurement is different so thus the information transfer. In this
section we discuss the flow of information as calculating the fidelity
for conclusive teleportation.

The fidelity ${\cal F}$ is defined by the overlap between the original
state $|\phi\rangle$ and the evolved state $\hat{\rho}$; ${\cal F} =
\langle \phi | \hat{\rho} | \phi \rangle$. When the quantum channel is
pure, a pure state is recovered at the receiving station after
performing one teleportation procedure.  The teleported pure state of
the density operator $\hat{\rho}_\alpha$ is dependent on the
measurement outcome, here, indexed $\alpha$, at the sending station.
After executing the teleportation protocol infinite times, the
ensemble of teleported quantum system is represented by a density
operator $\hat{\rho}=\sum_\alpha p_\alpha \hat{\rho}_\alpha$ where the
measurement bears the outcome indexed $\alpha$ with the probability
$p_\alpha$. The fidelity can thus be given by ${\cal F}= \sum_\alpha
p_\alpha \langle \phi | \hat{\rho}_\alpha | \phi \rangle$.  In quantum
teleportation, the original state $|\phi\rangle$ is unknown so that it
is necessary to average the fidelity over all possible unknown states.
The average fidelity is
\begin{equation}
\label{eq:fidleity}
\bar{\cal F} \equiv \frac{1}{V} \int d{\vec \Omega} \sum_\alpha
p_\alpha({\vec \Omega}) f_\alpha({\vec \Omega}) 
\end{equation}
where $f_\alpha({\vec \Omega}) = \langle \phi({\vec \Omega}) |
\hat{\rho}_\alpha | \phi({\vec \Omega}) \rangle$ and an unknown pure
states $|\phi({\vec \Omega})\rangle$ is parameterized by a real vector
${\vec \Omega}$ in the parameter space of volume $V$
\cite{Banaszek00,Horodecki96,Schack94}.

In conclusive teleportation, we know that faithful teleportation is
assured at the conclusive event.  Even though the inconclusive result
is not of use in quantum sense, the receiver can still try to recover
some information on the original unknown state.  To consider
conclusive and inconclusive events, we decompose further the
measurement operator (\ref{eq:last}) into general projectors such that
the new set of POVM operators is represented by
\begin{equation}
  \label{eq:fdmo}
    \hat{M}^\prime_\alpha = \lambda_\alpha^\prime |\tilde{\psi}^\prime_{\alpha}
    \rangle\langle \tilde{\psi}^\prime_{\alpha}|
\end{equation}
where
\begin{eqnarray}
  \label{eq:parastate}
  \lambda_\alpha^\prime &=& \left\{
    \begin{array}{ll}
      \lambda            & ~\mbox{for}~ \alpha \le d^2 \\
      1-\lambda/d a_j^2  & ~\mbox{for}~ \alpha = d^2 +(j-1)d + i
    \end{array} \right. \nonumber \\
  |\tilde{\psi}^\prime_\alpha\rangle &=& \left\{
    \begin{array}{ll}
      |\tilde{\psi}_\alpha\rangle  & ~\mbox{for}~ \alpha \le d^2 \\
      |ij\rangle  & ~\mbox{for}~ \alpha = d^2 +(j-1)d + i.
    \end{array} \right. \nonumber 
\end{eqnarray}
The average fidelity is now written as
\begin{eqnarray}
  \label{eq:split}
  \bar{\cal F}&=&\bar{\cal F}_{con}+\bar{\cal F}_{inc} \nonumber\\
  &=&\frac{1}{V}\int d{\vec \Omega} \Big[\sum^{d^2}_{\alpha=1}
  +\sum^{2d^2}_{\alpha=d^2+1}\Big] p_\alpha({\vec \Omega})
  f_\alpha({\vec \Omega}) 
\end{eqnarray}
where the first (second) sum indicates the information transferred in
conclusive (inconclusive) events. The average fidelity is maximized by
proper unitary operator $\hat{U}_\alpha$ according to the joint
measurement outcome for $\hat{M}'_\alpha$.  The resulting maximal
average fidelity is given by
\begin{equation}
  \label{eq:otaf}
  \bar{\cal F} = \lambda + \frac{1-\lambda}{d+1} +
  \frac{1}{d(d+1)}\left( \sum_{i=1}^d \sqrt{d a_i^2 -
  \lambda}\right)^2.
\end{equation}
Note that Eq.~(\ref{eq:otaf}) is obtained using the set of the POVM
operators $\{\hat{M}'_\alpha\}$. For the conclusive events the POVM
operators are pre-determined. On the other hand, the inconclusive
operators may vary to increase the fidelity. It was shown that the
fidelity of teleportation is maximized by orthogonal joint measurement
and unitary transformation \cite{Banaszek00}, which is consistent with
our choice of $\{\hat{M}'_\alpha\}$ for inconclusive events.

\subsection*{Information loss}
For the comparison with the standard teleportation we restrict, in
this section, our concern to two dimensional conclusive teleportation
where the parameter space of unknown states is the surface of a
three-dimensional sphere with its volume $V=4\pi$. When the quantum
channel is in the entangled state $|\psi\rangle_{23}= 2^{-1/2} [
\sqrt{1-\cos\theta_c} |11\rangle + \sqrt{1+\cos\theta_c} |22\rangle
]$, {$|\tilde{\psi}_\alpha^\prime\rangle_{12}$} in Eq.~(\ref{eq:fdmo})
are given by
\begin{eqnarray}
  \label{eq:jmoit}
  |\tilde{\psi}_1^\prime\rangle_{12} &=&
   N \left[\sqrt{1+\cos\theta_c}|11\rangle +
   \sqrt{1-\cos\theta_c}|22\rangle\right],  
   \nonumber \\
  |\tilde{\psi}_2^\prime\rangle_{12} &=&
   N \left[\sqrt{1+\cos\theta_c}|11\rangle -
   \sqrt{1-\cos\theta_c}|22\rangle\right], 
   \nonumber \\
  |\tilde{\psi}_3^\prime\rangle_{12} &=&
   N \left[\sqrt{1+\cos\theta_c}|21\rangle +
   \sqrt{1-\cos\theta_c}|12\rangle\right], 
   \nonumber \\
  |\tilde{\psi}_4^\prime\rangle_{12} &=&
   N \left[\sqrt{1+\cos\theta_c}|21\rangle -
   \sqrt{1-\cos\theta_c}|12\rangle\right],
\end{eqnarray}
where $N=(2-2\cos^2\theta_c)^{-1/2}$ and
$|\tilde{\psi}_\alpha^\prime\rangle_{12}=|ij\rangle$ for $\alpha > 4$.
The subscript indices $12$ have been added to emphasize that the
measurement is performed on particles 1 and 2. The non-orthogonality
of POVM operators is implied in the relative angle $\theta_c$ between
$|\tilde{\psi}_{1(3)}^\prime\rangle$ and
$|\tilde{\psi}_{2(4)}^\prime\rangle$.

The fidelity of each conclusive event is unity with the event
probability $\lambda/4$; $\bar{\cal F}_{con}=\lambda$.  For the
inconclusive event $\alpha$, the teleported state is represented by
the density operator $\hat{\rho}_\alpha = \frac{1}{2} (\openone +
\vec{m}_\alpha \cdot \vec{\sigma})$ where $\vec{m}_5=\hat{z}$,
$\vec{m}_6=-\hat{z}$, $\vec{m}_7=-\hat{z}$, and $\vec{m}_8=\hat{z}$,
are the four Bloch vectors. The average fidelity over the inconclusive
ensemble is
\begin{eqnarray}
  \label{eq:inconculsive}
  \bar{\cal F}_{inc}=\frac{1}{2}\left(1-\lambda\right)
  +&&\frac{1}{12}\Big[\left(a_1^2
    -\frac{\lambda}{2}\right)\left(m^z_5 -m^z_8\right) \nonumber \\
  && + \left(a_2^2
  -\frac{\lambda}{2}\right)\left(m^z_7-m^z_6\right)\Big]. 
\end{eqnarray} 
Because $\lambda \le 2~\mbox{min}\{a_1^2, a_2^2\}$ due to the
positivity of the measurement operators, Eq.~(\ref{eq:inconculsive})
is maximized to $\bar{\cal F}_{inc}=2 (1-\lambda)/3$ by applying
$\sigma_x$ on the receiver's particle for the outcomes $\alpha=7,8$.
Note that each inconclusive event has the fidelity 2/3 on average,
which is larger than the fidelity of 1/2 in any random guess process.
We have found that even the inconclusive events transfer some
information of unknown states to the receiving station.

The maximum average fidelity is finally obtained for a given parameter
$\lambda$ as
\begin{equation}
\label{eq:overall}
\bar{\cal F}=\frac{2}{3}\left(1+\frac{\lambda}{2}\right).
\end{equation}
It is interesting to note that the parameter $\lambda =
2~\mbox{min}\{a_1^2, a_2^2\}$ optimizes not only the probabilities for
conclusive events but also the average fidelity for conclusive
teleportation under the condition of identification
(\ref{eq:identify}). The average fidelity $\bar{\cal
  F}_o=\frac{2}{3}(1+2~\mbox{min}\{a_1^2, a_2^2\})$ for optimal
conclusive teleportation is clearly less than the average fidelity
$\bar{\cal F}_s$ for standard teleportation.  The optimal conclusive
teleportation has a smaller average fidelity than the standard
teleportation when the quantum channel is partially entangled.

\begin{figure}[htbp]
  \begin{center}
    \leavevmode
    \includegraphics[width=0.5\textwidth]{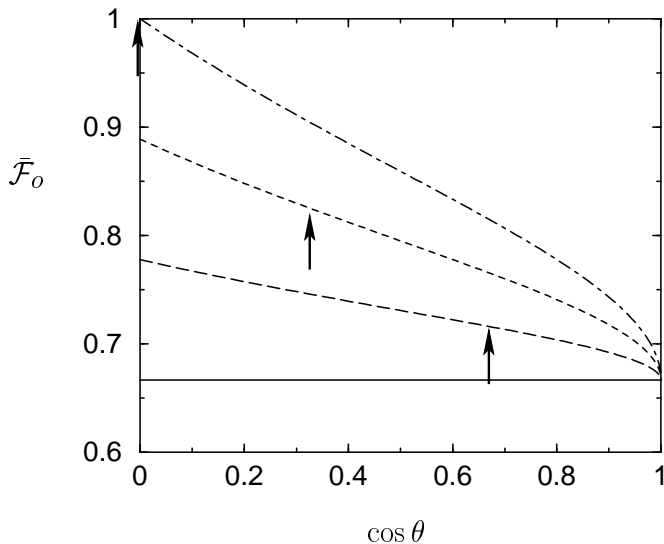}
  \end{center}
  \caption{The change of average optimum fidelity with regard to the relative
    angle of POVM operators for given entanglement of the quantum
    channel.  The degree of entanglement for the quantum channel is
    measured by the von-Neumann entropy ${\cal S}_a = 0$ (solid line),
    ${\cal S}_a = 0.19$ (long dashed), ${\cal S}_a = 0.55$ (dashed),
    and ${\cal S}_a = 1$ (dot dashed). The value of $\cos\theta=0$
    corresponds to standard teleportation. The arrows indicate the
    optimum fidelities of conclusive teleportation for given
    entanglement of the quantum channel.}
  \label{fig:fidelity}
\end{figure}

Why does the conclusive teleportation have a smaller average fidelity?
We will show that the information loss is caused by the
non-orthogonality in the joint POVM. To see this we release the
condition (\ref{eq:jmoit}) for conclusive identification and we vary
the relative angle $\theta_c \rightarrow \theta$.  Now, the new
measurement set represents also a POVM but does not necessarily give
us any conclusive result.  After some calculation, we find that, for a
given $\theta$, the new set of POVM operators leads to the maximal
average fidelity
\begin{equation}
  \label{eq:new}
  \bar{\cal F}=\frac{2}{3}\left(1+\frac{\lambda}{2}
  \frac{\sqrt{1-\cos^2\theta_c}} {\sqrt{1-\cos^2\theta}}\right),
\end{equation}
where the real parameter $\lambda$ is $0 \le \lambda \le
1-|\cos\theta|$. For each case of the new POVM set, the optimal
fidelity $\bar{\cal F}_o$ is obtained when $\lambda = 1-|\cos\theta|$.
Fig.~\ref{fig:fidelity} shows the dependence of the optimal fidelity
$\bar{\cal F}_o$ on the overlap, $\cos\theta$, of the POVM operators.
This clearly illustrates the dependence of teleported information on
the non-orthogonal nature of measurement operators. Four sets of data
are plotted with respect to the degree of channel entanglement.  The
degree of channel entanglement is measured by the von-Neumann entropy
${\cal S}_{a} \equiv \mbox{Tr}_a \hat{\rho}_a \log_2 \hat{\rho}_a$
where $\hat{\rho}_a$ is the reduced density operator for particle $a$.
When the overlap between the measurement bases is zero, {\it i.e.},
$\cos\theta=0$, the POVM set corresponds to the orthogonal
measurement. The arrows in Fig.~\ref{fig:fidelity} indicate the
fidelities of optimized conclusive teleportation for given quantum
channels. The dot-dashed line corresponds to the maximally entangled
channel and gives the unit fidelity for the orthogonal measurement of
$\cos\theta_c=0$ as it should in the standard teleportation. When the
non-orthogonality increases in the measurement, {\it i.e.},
$\cos\theta$ increases, the average fidelity decreases.  We can thus
say that the information of the unknown state is lost through the
non-orthogonal joint measurement in teleportation.

\section{Remarks}
\label{sec:sec4}

We have formulated the conclusive teleportation protocol utilizing the
joint POVM for an unknown state in the $d$-dimensional Hilbert space.
The general schemes are proposed on the identification of the
non-orthogonal states by POVM operators and the optimization of the
probabilities for the conclusive events. By the conclusive
teleportation one can teleport perfectly the unknown quantum state
with finite probability.  It is shown that some useful information for
teleportation can be extracted from an inconclusive event even when
the POVM is optimized.  The maximum fidelity for conclusive
teleportation has been found and the minimum resources have been
discussed.  The fidelity for conclusive teleportation is less than the
standard teleportation.  We attribute the reason for the loss of
information to the non-orthogonality of joint POVM operators.

\acknowledgements

This work has been supported by the BK21 Grant of the Korea Ministry
of Education.

\end{document}